\documentclass[sigconf, timestamp]{acmart}
\usepackage{textcomp}
\usepackage[shortcuts]{extdash}
\acmConference{SYSML 2018}{February 16, 2018}{Stanford University} 
\acmPrice{}
\acmISBN{}
\acmDOI{}
\setcopyright{rightsretained}
\newcommand{\Intel}{
\affiliation{%
	\institution{Intel}}}
\title{Intel\textregistered\ nGraph\texttrademark}
\subtitle{An Intermediate Representation, Compiler, and Executor for Deep Learning}
\author{Scott Cyphers}
\orcid{0000-0002-6661-2945}
\email{scott.cyphers@intel.com}
\author{Arjun K. Bansal}
\email{arjun.bansal@intel.com}
\author{Anahita Bhiwandiwalla}
\email{anahita.bhiwandiwalla@intel.com}
\author{Jayaram Bobba}
\email{jayaram.bobba@intel.com}
\author{Matthew Brookhart}
\email{matthew.i.brookhart@intel.com}
\author{Avijit Chakraborty}
\email{avijit.chakraborty@intel.com}
\author{Will Constable}
\email{will.h.constable@intel.com}
\Intel
\author{Christian Convey}
\email{christian.convey@intel.com}
\author{Leona Cook}
\email{leona.cook@intel.com}
\author{Omar Kanawi}
\email{omar.kanawi@intel.com}
\author{Robert Kimball}
\email{robert.kimball@intel.com}
\author{Jason Knight}
\email{jason.knight@intel.com}
\author{Nikolay Korovaiko}
\email{nikolay.korovaiko@intel.com}
\author{Varun Kumar}
\email{varun.v.kumar@intel.com}
\Intel
\author{Yixing Lao}
\email{yixing.lao@intel.com}
\author{Christopher R. Lishka}
\email{christopher.r.lishka@intel.com}
\author{Jaikrishnan Menon}
\email{jaikrishnan.menon@intel.com}
\author{Jennifer Myers}
\email{jennifer.myers@intel.com}
\author{Sandeep Aswath Narayana}
\orcid{0000-0002-8522-0891}
\email{sandeep.aswath.narayana@intel.com}
\author{Adam Procter}
\orcid{0000-0003-0005-8934}
\email{adam.m.procter@intel.com}
\author{Tristan J. Webb}
\email{tristan.webb@intel.com}
\Intel
\date{}	
\begin{document}
\begin{abstract}
The Deep Learning (DL) community sees many novel topologies published each year. Achieving high performance on each new topology remains challenging, as each requires some level of manual effort. This issue is compounded by the proliferation of frameworks and hardware platforms. The current approach, which we call ``direct optimization'', requires deep changes within each framework to improve the training performance for each hardware backend (CPUs, GPUs, FPGAs, ASICs) and requires $\mathcal{O}(fp)$ effort; where $f$ is the number of frameworks and $p$ is the number of platforms. While optimized kernels for deep-learning primitives are provided via libraries like Intel\textregistered\  Math Kernel Library for Deep Neural Networks (MKL-DNN), there are several compiler-inspired ways in which performance can be further optimized. Building on our experience creating neon (a fast deep learning library on GPUs), we developed Intel nGraph, a soon to be open-sourced C++ library to simplify the realization of optimized deep learning performance across frameworks and hardware platforms. Initially-supported frameworks include TensorFlow, MXNet, and Intel\textregistered\ neon framework. Initial backends are Intel Architecture CPUs (CPU), the Intel\textregistered\ Nervana Neural Network Processor\texttrademark\ (NNP), and NVIDIA GPUs. Currently supported compiler optimizations include efficient memory management and data layout abstraction. In this paper, we describe our overall architecture and its core components. In the future, we envision extending nGraph API support to a wider range of frameworks, hardware (including FPGAs and ASICs), and compiler optimizations (training versus inference optimizations, multi-node and multi-device scaling via efficient sub-graph partitioning, and HW-specific compounding of operations).
\end{abstract}
\maketitle
\renewcommand{\shortauthors}{Cyphers, Bansal, et al.}
\section{Introduction}
\begin{figure}
\includegraphics[scale=.5]{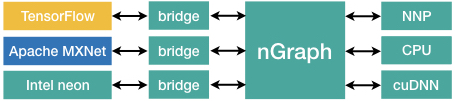}
\end{figure}

Deep learning frameworks are libraries that provide domain-specific languages for defining deep learning computations and APIs for managing data and executing computations. Backends' implementations are encapsulated so that the same computation can execute on multiple backends, such as CPUs and GPUs. We adopt the organization of compilers such as LLVM\cite{LLVM:CGO04} by converting framework-specific computation definitions into a framework-independent intermediate representation (IR) that we compile into a form that can execute on the backend. An nGraph \emph{framework bridge} acts as a framework backend. Each nGraph backend has a \emph{transformer} that compiles or interprets the IR and provides an allocation and execution API that the framework bridges use to implement the framework's API. A key difference between compilers for languages like C++ and compilers for deep learning frameworks is that with deep learning, the data being operated on is large and variable-sized, but highly amenable to parallelization.

\subsection{Related Work}
Google's accelerated linear algebra (XLA)\cite{XLA} compiler acts as an experimental backend for TensorFlow\cite{tensorflow2015-whitepaper}. Unlike nGraph, the XLA project has made no public comments about support of frameworks other than TensorFlow. 

The DMLC group announced the NNVM\cite{nnvm} project as a light-weight graph optimization library for deep learning and later announced TVM\cite{tvm} as an ahead-of-time compiler that supports multiple hardware platforms and interoperates with NNVM. NNVM leaves operator set unspecified, which makes different frontends and backends incompatible. nGraph, XLA, and LLVM use a fixed, but extensible, IR operation set.

DLVM \cite{dlvm} is a project of the University of Illinois Urbana\-/Champaign. DLVM proposes an LLVM inspired modular IR with full control flow and a side-effect free representation. It remains to be seen if the more flexible IR is capable of supporting the performance optimizations enabled by simpler data flow graph IRs like those of nGraph, XLA, and NNVM.

ONNX \cite{onnx} is a recent cross-industry effort, which we participate in, to standardize an IR for inference. The nGraph IR has a richer feature set, including support for training and a rich set of optimization passes and backends for execution. We will aim for 
ONNX interoperability.

The activity in the space of deep learning compilers and IRs highlights their need and we look forward to a healthy exchange of ideas as the field moves forward on these complementary efforts. We believe nGraph can interoperate with developing standards through framework bridges and nGraph backends.

\section{Intermediate Representation}
An nGraph IR is a directed acyclic graph of stateless operation nodes. Each node has zero or more inputs and zero or more outputs. Nodes may have additional constant attributes that affect their behavior, such as which axes to sum over. The inputs and attributes of a node determine the shape and element types of the outputs.

Nodes operate on multi-dimensional arrays, called tensors. Most frameworks associate semantics with particular axis positions. For example, images might be stored in tensors ordered by mini-batch size, channels, height and width. The framework-specific ordering is usually a reflection of op implementation and tensor element layout. When dealing with video or high-dimensional time series datasets, rank restrictions from framework ops require explicit tensor reshaping and axis reordering. With the exception of tensors directly accessible to framework users, nGraph, \emph{does not} have a fixed relationship between axis order and tensor element layout.

\section{Framework Bridges}
Frameworks use a framework-specific symbolic representation of their computations, called a \emph{computational graph}. Backends use the graph to interpret or compile computations. The graph is also used in the implementation of some form of the autodiff algorithm for the computation of derivatives, either by computing the derivative directly on the graph, or by computing the graph for a derivative computation from an existing graph. Framework bridges belonging to the nGraph library use the graph to construct the nGraph IR.

Apache MXNet\cite{mxnet} is a core C++ library and a C API for interacting with several frontend languages. Machine learning models may be defined imperatively through Gluon or symbolically through the standard MXNet frontend. Models' operations are represented as nodes in the NNVM graph. The MXNet-nGraph bridge translates the NNVM inference graph into nGraph IR; it selects the largest possible computation for the respective backend and uses autodiff on the nGraph IR for the derivative. Compiled nGraph functions can then interface with the standard MXNet execution engines.

TensorFlow's\cite{tensorflow2015-whitepaper} XLA framework enables compilation and execution of TensorFlow graphs on novel hardware such as NNP. During the execution of a computation via XLA, an HLO IR of the TensorFlow computation is sent to a  device such as a CPU, GPU or novel hardware.  Our bridge plugin registers itself as a new XLA device, maps HLO IR to nGraph IR, and returns a compiled function. During the execution of the TensorFlow computation, the function is invoked by TensorFlow on the input data and the resulting nGraph output is returned.

For neon, we are creating a Python binding for the nGraph API, which we hope to also use with other Python-based frameworks.

\section{Transformers}
The IR generated by the nGraph library is passed to a transformer for the generation of code optimized specifically for the selected backend. These newly-optimized backends provide facilities for pattern matching, liveness analysis, memory management, and the combining of tensor-element layout and shape management with backend kernel selection. 

The CPU transformer makes use of MKL-DNN, which produces optimized sequences of calls to highly optimized kernels. Optimizations provided by MKL-DNN are at a finer granularity than those provided by nGraph. The CPU transformer will also be used by other transformers for sub-graphs that use operations not supported by their backend.

Intel's NNP processor is tailored for deep learning workloads. Its transformer lets us make the fullest use of the hardware, falling back on the CPU transformer for unsupported operations.

The cuDNN\cite{cudnn} transformer dynamically generates code that links to the NVIDIA CUDA Deep Neural Network (cuDNN) library for common kernels, such as convolution or softmax. The nGraph library then compiles portions of the graph into LLVM IR representation, and uses the PTX-emitting backend of LLVM to generate the GPU assembly language PTX.

nGraph will natively support collective communication primitives (AllReduce, Gather, Broadcast), as well as point-to-point primitives as core graph ops.  Transformers will generate the corresponding communication library calls. Transformers will support vanilla MPI, or provide optimized communication methods that impose restrictions. An example of such a restriction would be: operate only across a homogeneous cluster of a certain topology.
\bibliographystyle{ACM-Reference-Format}
\bibliography{ngraph}


\begin{thebibliography}{9}


\ifx \showCODEN    \undefined \def \showCODEN     #1{\unskip}     \fi
\ifx \showDOI      \undefined \def \showDOI       #1{#1}\fi
\ifx \showISBNx    \undefined \def \showISBNx     #1{\unskip}     \fi
\ifx \showISBNxiii \undefined \def \showISBNxiii  #1{\unskip}     \fi
\ifx \showISSN     \undefined \def \showISSN      #1{\unskip}     \fi
\ifx \showLCCN     \undefined \def \showLCCN      #1{\unskip}     \fi
\ifx \shownote     \undefined \def \shownote      #1{#1}          \fi
\ifx \showarticletitle \undefined \def \showarticletitle #1{#1}   \fi
\ifx \showURL      \undefined \def \showURL       {\relax}        \fi
\providecommand\bibfield[2]{#2}
\providecommand\bibinfo[2]{#2}
\providecommand\natexlab[1]{#1}
\providecommand\showeprint[2][]{arXiv:#2}

\bibitem[\protect\citeauthoryear{??}{mxn}{2017}]%
        {mxnet}
 \bibinfo{year}{2017}\natexlab{}.
\newblock \bibinfo{title}{Apache MXNet}.
\newblock   (\bibinfo{year}{2017}).
\newblock
\urldef\tempurl%
\url{https://mxnet.apache.org/}
\showURL{%
Retrieved January 4, 2018 from \tempurl}


\bibitem[\protect\citeauthoryear{??}{onn}{2017}]%
        {onnx}
 \bibinfo{year}{2017}\natexlab{}.
\newblock \bibinfo{title}{ONNX}.
\newblock   (\bibinfo{year}{2017}).
\newblock
\urldef\tempurl%
\url{http://onnx.ai}
\showURL{%
Retrieved January 4, 2018 from \tempurl}


\bibitem[\protect\citeauthoryear{??}{cud}{2018}]%
        {cudnn}
 \bibinfo{year}{2018}\natexlab{}.
\newblock \bibinfo{title}{NVIDIA cuDNN}.
\newblock   (\bibinfo{year}{2018}).
\newblock
\urldef\tempurl%
\url{https://developer.nvidia.com/cudnn}
\showURL{%
Retrieved January 4, 2018 from \tempurl}


\bibitem[\protect\citeauthoryear{Abadi, Agarwal, Barham, Brevdo, Chen, Citro,
  Corrado, Davis, Dean, Devin, Ghemawat, Goodfellow, Harp, Irving, Isard, Jia,
  Jozefowicz, Kaiser, Kudlur, Levenberg, Man\'{e}, Monga, Moore, Murray, Olah,
  Schuster, Shlens, Steiner, Sutskever, Talwar, Tucker, Vanhoucke, Vasudevan,
  Vi\'{e}gas, Vinyals, Warden, Wattenberg, Wicke, Yu, and Zheng}{Abadi
  et~al\mbox{.}}{2015}]%
        {tensorflow2015-whitepaper}
\bibfield{author}{\bibinfo{person}{Mart\'{\i}n Abadi}, \bibinfo{person}{Ashish
  Agarwal}, \bibinfo{person}{Paul Barham}, \bibinfo{person}{Eugene Brevdo},
  \bibinfo{person}{Zhifeng Chen}, \bibinfo{person}{Craig Citro},
  \bibinfo{person}{Greg~S. Corrado}, \bibinfo{person}{Andy Davis},
  \bibinfo{person}{Jeffrey Dean}, \bibinfo{person}{Matthieu Devin},
  \bibinfo{person}{Sanjay Ghemawat}, \bibinfo{person}{Ian Goodfellow},
  \bibinfo{person}{Andrew Harp}, \bibinfo{person}{Geoffrey Irving},
  \bibinfo{person}{Michael Isard}, \bibinfo{person}{Yangqing Jia},
  \bibinfo{person}{Rafal Jozefowicz}, \bibinfo{person}{Lukasz Kaiser},
  \bibinfo{person}{Manjunath Kudlur}, \bibinfo{person}{Josh Levenberg},
  \bibinfo{person}{Dan Man\'{e}}, \bibinfo{person}{Rajat Monga},
  \bibinfo{person}{Sherry Moore}, \bibinfo{person}{Derek Murray},
  \bibinfo{person}{Chris Olah}, \bibinfo{person}{Mike Schuster},
  \bibinfo{person}{Jonathon Shlens}, \bibinfo{person}{Benoit Steiner},
  \bibinfo{person}{Ilya Sutskever}, \bibinfo{person}{Kunal Talwar},
  \bibinfo{person}{Paul Tucker}, \bibinfo{person}{Vincent Vanhoucke},
  \bibinfo{person}{Vijay Vasudevan}, \bibinfo{person}{Fernanda Vi\'{e}gas},
  \bibinfo{person}{Oriol Vinyals}, \bibinfo{person}{Pete Warden},
  \bibinfo{person}{Martin Wattenberg}, \bibinfo{person}{Martin Wicke},
  \bibinfo{person}{Yuan Yu}, {and} \bibinfo{person}{Xiaoqiang Zheng}.}
  \bibinfo{year}{2015}\natexlab{}.
\newblock \bibinfo{title}{{TensorFlow}: Large-Scale Machine Learning on
  Heterogeneous Systems}.
\newblock   (\bibinfo{year}{2015}).
\newblock
\urldef\tempurl%
\url{https://www.tensorflow.org/}
\showURL{%
\tempurl}
\newblock
\shownote{Software available from tensorflow.org.}


\bibitem[\protect\citeauthoryear{Chen, Moreau, Jiang, and Shen}{Chen
  et~al\mbox{.}}{2017}]%
        {tvm}
\bibfield{author}{\bibinfo{person}{Tianqi Chen}, \bibinfo{person}{Thierry
  Moreau}, \bibinfo{person}{Ziheng Jiang}, {and} \bibinfo{person}{Haichen
  Shen}.} \bibinfo{year}{2017}\natexlab{}.
\newblock \bibinfo{title}{TVM: An End to End IR Stack for Deploying Deep
  Learning Workloads on Hardware Platforms}.
\newblock   (\bibinfo{date}{Aug} \bibinfo{year}{2017}).
\newblock
\urldef\tempurl%
\url{http://tvmlang.org/2017/08/17/tvm-release-announcement.html}
\showURL{%
\tempurl}


\bibitem[\protect\citeauthoryear{Google}{Google}{2017}]%
        {XLA}
\bibfield{author}{\bibinfo{person}{Google}.} \bibinfo{year}{2017}\natexlab{}.
\newblock \bibinfo{title}{XLA Overview}.
\newblock   (\bibinfo{year}{2017}).
\newblock
\urldef\tempurl%
\url{https://www.tensorflow.org/performance/xla/}
\showURL{%
Retrieved January 5, 2018 from \tempurl}


\bibitem[\protect\citeauthoryear{Lattner and Adve}{Lattner and Adve}{2004}]%
        {LLVM:CGO04}
\bibfield{author}{\bibinfo{person}{Chris Lattner} {and} \bibinfo{person}{Vikram
  Adve}.} \bibinfo{year}{2004}\natexlab{}.
\newblock \showarticletitle{{LLVM: A Compilation Framework for Lifelong Program
  Analysis \& Transformation}}. In \bibinfo{booktitle}{\emph{{Proceedings of
  the 2004 International Symposium on Code Generation and Optimization
  (CGO'04)}}}. \bibinfo{address}{Palo Alto, California}.
\newblock


\bibitem[\protect\citeauthoryear{team}{team}{2017}]%
        {nnvm}
\bibfield{author}{\bibinfo{person}{Amazon Web Service~AI team}.}
  \bibinfo{year}{2017}\natexlab{}.
\newblock   (\bibinfo{date}{Oct} \bibinfo{year}{2017}).
\newblock
\urldef\tempurl%
\url{http://tvmlang.org/2017/10/06/nnvm-compiler-announcement.html}
\showURL{%
\tempurl}


\bibitem[\protect\citeauthoryear{Wei, Adve, and Schwartz}{Wei
  et~al\mbox{.}}{2017}]%
        {dlvm}
\bibfield{author}{\bibinfo{person}{Richard Wei}, \bibinfo{person}{Vikram~S.
  Adve}, {and} \bibinfo{person}{Lane Schwartz}.}
  \bibinfo{year}{2017}\natexlab{}.
\newblock \showarticletitle{{DLVM:} {A} modern compiler infrastructure for deep
  learning systems}.
\newblock \bibinfo{journal}{\emph{CoRR}}  \bibinfo{volume}{abs/1711.03016}
  (\bibinfo{year}{2017}).
\newblock
\showeprint[arxiv]{1711.03016}
\urldef\tempurl%
\url{http://arxiv.org/abs/1711.03016}
\showURL{%
\tempurl}


\end{thebibliography}
\end{document}